\documentstyle[aps,amsfonts,amssymb,preprint]{revtex}
\tighten
\draft

\newcommand{\be}{\begin{equation}}
\newcommand{\ee}{\end{equation}}

\begin{document}
\preprint{OUTP-00-23-P:gr-qc/0005108}
\date{\today}

\title{Divergence of a quantum thermal state on Kerr
space-time}

\author{Adrian C. Ottewill} 
\address{Department of Mathematical Physics, University College Dublin,
\\ Belfield, Dublin 4, Ireland\\
email: ottewill@relativity.ucd.ie}

\author{Elizabeth Winstanley}
\address{Department of Theoretical Physics, University of Oxford,\\ 
1 Keble Road, Oxford OX1 3NP, UK\\
email: elizabeth.winstanley@oriel.ox.ac.uk}

\maketitle
\begin{abstract}
We present a simple proof, using the conservation equations,
that any quantum stress tensor on Kerr space-time which is 
isotropic in a frame which rotates rigidly with the
angular velocity of the event horizon must be
divergent at the velocity of light surface.
We comment on our result
in the light of the absence of a `true
Hartle-Hawking' vacuum for Kerr. 
\end{abstract}

One of the fundamental results of quantum field theory
in Kerr space-time is a theorem of Kay and Wald~\cite{kw},
that there does not exist a Hadamard state which respects
the symmetries of the space-time and is regular everywhere 
outside and on the event horizon.
This means that there is no `true Hartle-Hawking' (HH) vacuum
on Kerr space-time.  
The HH vacuum on Schwarzschild black holes
has been extensively studied, since it is the state
which lends itself most readily to numerical computations,
due to its regularity and high degree of symmetry.
For the same reason the construction of a state on Kerr
space-time with most (but not all) of the properties of the HH
state remains an important open question.

The construction of quantum states on Kerr space-time
is a delicate matter due to the presence of the
super-radiant modes (see~\cite{ow} for details of 
this procedure). 
There is a consensus in the literature that the (past)
Boulware vacuum $|B^-\rangle $ is defined by 
taking a basis of modes which are positive
frequency with respect to the Killing vector
$\partial /\partial t$ at ${\mathfrak {I}}^{-}$
and with respect to the Killing vector $\partial /\partial t +
\Omega_H \partial /\partial \phi$ (where $\Omega_H$ is the angular
velocity of the horizon) at ${\mathfrak {H}}^{-}$.
This state corresponds to an absence of particles
coming up from ${\mathfrak {H}}^{-}$ or in from
${\mathfrak {I}}^{-}$, and contains, at
${\mathfrak {I}}^{+}$, an outward flux of particles
due to the Unruh-Starobinskii effect (spontaneous emission in
superradiant modes).
There are two attempts in the literature to define 
a state analogous to the HH 
state~\cite{ft,cch}, one due to Frolov and Thorne~\cite{ft}
and the other due to  Candelas, Chrzanowski and
Howard~\cite{cch} which we shall denote by
$|FT\rangle $ and $|CCH \rangle$, respectively.
We refer the reader to~\cite{ow} for details of
the construction of these states, which are 
not important here.  
Both these candidate states are thermal in nature
(although the energy with respect to which the
modes are thermalized differ), but they
have different symmetry and regularity properties.
The state $|FT\rangle $ is invariant under simultaneous
$t$, $\phi $ reversal, while $|CCH \rangle $ is not.

Our interest in this letter is in the difference
in expectation values of the quantum stress tensor
in the Boulware and candidate HH vacua.
By considering such a difference any renormalization terms
cancel and we may deal effectively deal with the bare operators.
We shall denote this tensor as 
$\langle {\hat {T}}_{\mu \nu } \rangle _{HH-B}$ regardless of
which candidate HH vacuum we are considering,
since our result in this paper is quite general and
does not depend on the details of the construction
of the thermal states.
In~\cite{ft}, $\langle {\hat {T}}_{\mu \nu } \rangle _{HH-B}$ 
represents a precisely
thermal atmosphere of quanta at the Hawking temperature
which rotates rigidly
with the same angular velocity as the event horizon,
$\Omega _{H}$. 
Therefore, $\langle {\hat {T}}_{\mu \nu } \rangle _{HH-B}$ should be 
isotropic in a frame which also rotates rigidly
with angular velocity $\Omega _{H}$.
This is in agreement with the calculation of~\cite{cch}, 
where the difference in expectation values
of the stress tensor in the Boulware $|B^{-}\rangle $
and $|CCH\rangle $ states was calculated at the event 
horizon for an electromagnetic field. 
They found that this tensor was isotropic in the Carter
tetrad~\cite{carter}.
Due to the rotation of the black hole, this is the
same, at the event horizon, as the tensor being
isotropic in a tetrad which is rigidly rotating with 
the same angular velocity as the black hole.
The calculation in~\cite{cch} is valid only
close to the event horizon, and does not provide
any information about the isotropy (or otherwise)
of $\langle {\hat {T}}_{\mu \nu } \rangle _{HH-B}$ 
away from the horizon.

Frolov and Thorne~\cite{ft} then used an argument based
on the details of the quantum field modes to show that
this stress tensor will fail to be regular at the 
velocity of light surface ${\cal {S}}_{L}$.
This is the surface, where, in order to co-rotate 
with the event horizon, an observer must travel
at the speed of light.
We shall now show that it is an elementary consequence
of the conservation equations that
$\langle {\hat {T}}_{\mu \nu } \rangle _{HH-B}$, 
with the assumed isotropy property,
will fail to be regular at ${\cal {S}}_{L}$.
In addition, we will also find that 
$\langle {\hat {T}}_{\mu \nu } \rangle _{HH-B}$ has
the same, divergent, form at the event horizon as found
by~\cite{cch}.
Our metric has signature $(-+++)$ and we
use units in which $G=c=\hbar =k_{b}=1$ throughout.
Greek letters will denote co-ordinate components, 
while bracketed Roman letters denote tetrad components.

Firstly, we begin by writing the Kerr metric
in the unconventional form~\cite{ft}
\be
ds^{2} =
- \alpha ^{2} dt^{2} 
+ \rho ^{2} \Delta ^{-1} dr^{2} 
+ \rho ^{2} \, d\theta ^{2}
+ {\tilde {\Omega }}^{2} (d\phi - \Omega \, dt)^{2},
\label{eq:metric}
\ee
where $\rho ^{2}=r^{2} + a^{2} \cos ^{2} \theta $,
$\Delta = r^{2} - 2Mr + a^{2}$, with $M$ the mass of the
black hole and $a$ the angular momentum per unit mass, 
as viewed from infinity.
The other functions appearing in the metric (\ref{eq:metric})
are given by:
\begin{eqnarray}
\alpha ^{2} & = &
\frac {\Delta \rho ^{2}}{\left( r^{2} + a^{2} \right) ^{2}
- \Delta a^{2} \sin ^{2} \theta } ,
\nonumber 
\\
\Omega & = & 
\frac {2Mra}{\left( r^{2} + a^{2} \right) ^{2}
- \Delta a^{2} \sin ^{2} \theta } ,
\nonumber 
\\
{\tilde {\Omega }}^{2} & = & 
\frac {1}{\rho ^{2}} \left[ \left( r^{2} + a^{2} \right) ^{2}
- \Delta a^{2} \sin ^{2} \theta \right] \sin ^{2} \theta .
\end{eqnarray} 
Here $\alpha $ is the lapse function, vanishing on the event 
horizon $r=r_{H}$ (when $\Delta =0$), 
and $\Omega $ is the angular velocity of a 
locally non-rotating observer (LNRO), which is equal to the
angular velocity $\Omega _{H}$ of the event horizon when $r=r_{H}$.
An observer who is rotating with angular velocity $\Omega _{H}$
has the Lorentz factor $\gamma $ relative to a LNRO at the 
same values of $r$ and $\theta $, where
\be
\gamma = \left( 1-v^{2} \right) ^{-\frac {1}{2}},
\qquad
v= \alpha ^{-1} \left( \Omega _{H} -\Omega  \right) 
{\tilde {\Omega }}.
\label{eq:lorentz}
\ee
At the event horizon, an LNRO has angular velocity $\Omega _{H}$
and in this case $\gamma =1$, whilst at 
${\cal {S}}_{L}$, $v\rightarrow 1$ and 
$\gamma \rightarrow \infty $, as expected.

An orthonormal frame which rotates rigidly with angular
velocity $\Omega _{H}$ has basis 1-forms given by:
\begin{eqnarray}
e_{(t)i}dx^{i}
& = & \alpha ^{-1}\gamma \left[ \alpha ^{2} dt 
-{\tilde {\Omega }}^2
(\Omega _{H} -\Omega )
(d\phi -\Omega  \, dt) \right] ,
\nonumber \\
e_{(r)i}dx^{i} & = &
(\rho^2 \Delta ^{-1})^{\frac {1}{2}} dr ,
\nonumber \\
e_{(\theta )i}dx^{i} & = & \rho \, d\theta , 
\nonumber \\
e_{(\phi )i}dx^{i} & = & 
{\tilde {\Omega }}\gamma (d\phi -\Omega _{H} \, dt).
\label{eq:tetrad}
\end{eqnarray}
We shall consider a stress tensor 
$\langle {\hat {T}}_{\mu \nu } \rangle _{HH-B}$ which is isotropic
in this frame, so that with respect to the basis of 1-forms
(\ref{eq:tetrad}) the tetrad components are:
\be
\langle {\hat {T}}^{(a)}_{(b)} \rangle _{HH-B}
=f(r,\theta ) \, {\mathrm {diag}} \left\{ -3,1,1,1 \right\} ,
\ee
where we have used the Killing vector symmetries of the geometry
to assume that $\langle {\hat {T}}_{\mu \nu } \rangle _{HH-B}$ 
does not depend on either $t$ or $\phi $~\cite{ow}.
In order to find the unknown function $f(r,\theta )$, we solve
the conservation equations for 
$\langle {\hat {T}}_{\mu \nu } \rangle _{HH-B}$.
The simplest way to do this, which avoids the calculation 
of Ricci rotation coefficients for the tetrad~(\ref{eq:tetrad}),
is to convert the tetrad components back to Boyer-Lindquist 
co-ordinate components and then solve the conservation equations
in the form~\cite{ow}:
\be
\partial_{\nu } \left( \langle
{\hat {T}}_{\mu }^{\nu } \rangle _{HH-B} {\sqrt {-g}} 
\right) 
= \frac {1}{2} {\sqrt {-g}} \left( \partial _{\mu }
g_{\lambda \sigma }\right) 
\langle {\hat {T}}^{\lambda \sigma }
\rangle _{HH-B} .
\ee
The $\mu =t$ and $\mu =\phi $ equations are trivial, and 
the $\mu =r$ and $\mu =\theta $ equations give, respectively,
\begin{eqnarray}
\partial _{r} \left( f(r,\theta ) \rho ^{2} 
\sin \theta \right) & = &
\frac {1}{2} f(r,\theta ) \rho ^{2} \sin \theta \, 
\partial _{r} \left( \log | \alpha ^{-8}
\gamma ^{8} \rho ^{4} \sin ^{2} \theta | \right)
\nonumber \\
\partial _{\theta } \left( f(r,\theta ) \rho ^{2} \sin \theta 
\right)
& = &
\frac {1}{2} f(r,\theta ) \rho ^{2} \sin \theta \, 
\partial _{\theta }
\left( \log |\alpha ^{-8}\gamma ^{8} \rho ^{4} 
\sin ^{2}\theta | \right).
\end{eqnarray}
These two equations are compatible and determine 
$f(r,\theta )$ up to an arbitrary constant.  
The result is 
\be
f(r,\theta )=k\alpha ^{-4}\gamma ^{4}
\ee
where $k$ is an arbitrary constant.  
This implies that the tetrad components of 
$\langle {\hat {T}}_{\mu \nu } \rangle _{HH-B}$ diverge
as $\Delta ^{-2}$ as the event horizon is approached,
which is the behaviour found in~\cite{cch}.

In order to consider the regularity (or otherwise) of 
$\langle {\hat {T}}_{\mu \nu } \rangle _{HH-B}$
at the event horizon or ${\cal {S}}_{L}$, we must first convert the
tetrad components to Boyer-Lindquist components, since
the tetrad (\ref{eq:tetrad}) is not regular either at 
the event horizon or ${\cal {S}}_{L}$.
The non-zero components of 
$\langle {\hat {T}}_{\mu \nu } \rangle _{HH-B}$ are:
\begin{eqnarray}
\langle {\hat {T}}_{tt} \rangle _{HH-B} & = & 
\left[ 3\alpha ^{2} + {\tilde {\Omega }}^{2} \left(
4\gamma ^{2} \Omega _{H}^{2} - 3\Omega ^{2} \right)
\right] f(r,\theta ) ,
\nonumber
\\
\langle {\hat {T}}_{t\phi } \rangle _{HH-B} & = &
- {\tilde {\Omega }}^{2} \left(
4\gamma ^{2} \Omega _{H} - 3 \Omega \right)
f(r,\theta ) , 
\nonumber 
\\
\langle {\hat {T}}_{\phi \phi } \rangle _{HH-B} & = & 
{\tilde {\Omega }}^{2} \left( 4 \gamma ^{2} - 3 \right)
f(r, \theta ) ,
\nonumber
\\
\langle {\hat {T}}_{rr} \rangle _{HH-B} & = & 
\rho ^{2} \Delta ^{-1} f(r,\theta )  ,
\nonumber
\\
\langle {\hat {T}}_{\theta \theta } \rangle _{HH-B} & = & 
\rho ^{2} f(r,\theta ) .
\label{eq:blcomps}
\end{eqnarray}
Using these components, the regularity of 
$\langle {\hat {T}}_{\mu \nu } \rangle _{HH-B}$ on the event horizon
can be considered by first transforming to Kruskal co-ordinates,
as in~\cite{ow}.
Considering the Kruskal components shows that 
$\langle {\hat {T}}_{\mu \nu } \rangle _{HH-B}$ is not regular 
on either the past or the future event horizon, due to the
$\Delta ^{-2}$ divergence of $f$ as the horizon is approached.
This is precisely the behaviour expected of the Boulware vacuum
close to the event horizon, and does not preclude the 
possibility that a candidate HH state may be regular
on some section of the event horizon.
In~\cite{ow} it was argued that, of the two candidates for
the analogue of the HH vacuum in Kerr, $|FT \rangle $ is regular
only at the pole of the event horizon, and $|CCH \rangle $
is regular on the future (but not on the past) event horizon.
Both these cases are compatible with our results here, provided
that the divergences (where they exist) are of lower order
than the $\Delta ^{-2}$ expected for the Boulware vacuum.

The Boyer-Lindquist co-ordinate system is regular at ${\cal {S}}_{L}$,
and so (\ref{eq:blcomps}) reveals that the components of 
$\langle {\hat {T}}_{\mu \nu } \rangle _{HH-B}$
diverge at least as fast 
as $\gamma ^{4}$ as the velocity of light surface
is approached and $\gamma \rightarrow \infty $.
Thus, if $\langle B^{-}|{\hat {T}}_{\mu \nu}|B^{-}\rangle _{ren}$ 
is regular at the velocity of light surface, 
then the expectation value of the stress tensor 
in the HH vacuum diverges there.  
This is certainly the case for slowly 
rotating black holes when ${\cal {S}}_{L}$
is far from the horizon and  
$\langle B^{-}|{\hat {T}}_{\mu \nu}|B^{-}\rangle _{ren}$ 
is given by the Unruh-Starobinskii effect.
In fact, the Boulware vacuum is expected to be 
regular everywhere away from the event horizon 
(where it diverges).
We conclude that {\em {if}} we have a state
$|H\rangle $ such that $\langle {\hat {T}}_{\mu \nu } \rangle _{HH-B}$
is isotropic in the tetrad (\ref{eq:tetrad}), then
the state $|H\rangle $ will not be regular
on ${\cal {S}}_{L}$.

Thus we have shown that a stress tensor which is isotropic
with respect to a frame which rotates rigidly with the 
angular velocity of the event horizon must diverge as 
$\Delta ^{-2}$ as the event horizon is approached, and must
also fail to be regular at ${\cal {S}}_{L}$.
The question is therefore, is this isotropy condition likely
to be satisfied by a physical stress tensor?
The answer is far from obvious as the peak of the thermal
spectrum at the Hawking temperature is at a wavelength 
comparable to the radius of curvature of the space-time near the
horizon.
In the absence of a full numerical calculation in Kerr, we 
note that the corresponding property of 
$\langle {\hat {T}}_{\mu \nu } \rangle _{HH-B}$
for Schwarzschild black holes was conjectured in~\cite{cf}, 
and subsequently shown to hold to a good 
approximation in~\cite{jmo}.  
The authors of~\cite{ft} used the thermal properties of
their state $|FT \rangle $ to show that the corresponding
stress  tensor $\langle {\hat {T}}_{\mu \nu } \rangle _{HH-B}$
is isotropic in the rigidly rotating tetrad, although
in~\cite{ow} we have argued that, despite its attractive
symmetry properties (in particular, simultaneous $t$, $\phi $
reversal invariance) this state is fundamentally flawed.
The alternative state $|CCH \rangle $, which is not invariant
under simultaneous $t$, $\phi $ reversal, is more workable,
and satisfies the required isotropy property, at least near the 
event horizon~\cite{cch}.
Note that, in contrast to the situation in Schwarzschild, 
the Boulware vacuum in Kerr is not invariant under
simultaneous $t$,  $\phi $ reversal.
Therefore, we would expect any analogue HH state
such that $\langle {\hat {T}}_{\mu \nu } \rangle _{HH-B}$
has our conjectured isotropy must also not be $t$, $\phi $
reversal invariant.

It should be stressed that it is crucial to our
analysis that the stress tensor
$\langle {\hat {T}}_{\mu \nu } \rangle _{HH-B}$ 
is isotropic in the rigidly rotating frame~(\ref{eq:tetrad})
rather than any other tetrad.
For example, a stress tensor which is isotropic with respect to
the Carter tetrad~\cite{carter} everywhere
outside the event horizon will also display the
divergence near the event horizon we have exhibited here,
but will be regular elsewhere.
This would be anticipated from the fact that the 
Carter tetrad rotates with an angular velocity which,
although not the same as the angular velocity of the LNROs,
decreases as we move away from the event horizon.
Therefore, observers who are rotating with the same 
angular velocity as the Carter tetrad will always
have a finite Lorentz factor relative to the LNROs (compare
(\ref{eq:lorentz})).
The point of~\cite{ft} is that only observers who
have angular velocity $\Omega _{H}$ (whatever their
position outside the event horizon) will see an isotropic
thermal distribution of particles, which is the reason 
for our conjectured isotropy condition here.
The argument in~\cite{ft} uses the thermal
properties of their state $|FT \rangle $, and
does not apply to the other candidate HH state,
$|CCH \rangle $, due to the different thermalization
of the modes~\cite{ow}.
We hope to return to this question subsequently, by 
examining the isotropy (or otherwise) of these
candidate HH vacua using numerical calculations.

In the light of the Kay-Wald theorem~\cite{kw}, which states
that there is no true HH state on Kerr spacetime,
our result shows that, if it is possible to define
a state on Kerr which has most of the properties of the 
HH state, then either the stress tensor 
$\langle {\hat {T}}_{\mu \nu } \rangle _{HH-B}$  corresponding
to this state will fail to be isotropic in the rigidly rotating
tetrad, or the state will cease to be regular on ${\cal {S}}_{L}$.
Furthermore, we have shown that the divergence at ${\cal {S}}_{L}$ 
is an elementary consequence of the conservation equations.

The work of E.W. is supported by a fellowship from Oriel College, 
Oxford, and she would like to thank the Department of Physics,
University of Newcastle, Newcastle-upon-Tyne, for hospitality
during the completion of this work.

\end{document}